\documentclass[prl,showpacs,twocolumn,superscriptaddress]{revtex4}
\usepackage{graphicx}
\usepackage{color}
\usepackage{textcomp}
\usepackage{amsmath}
\usepackage{amssymb}

\definecolor{red}{rgb}{1,0.2706,0}
\definecolor{green}{rgb}{0.2706,1,0}
\definecolor{blue}{rgb}{0.2706,0,1}

\newcommand{\un}[1]{\,\text{#1}}
\newcommand{\bper}{\ensuremath{B_\perp}}
\newcommand{\boffs}{\ensuremath{B_\text{offset}}}

\begin{document}
\title{Kondo effect in a one-electron double quantum dot: Oscillations of the
Kondo current in a weak magnetic field}

\author{D.\,M.~Schr\"oer}
\email{schroeer@lmu.de}
\affiliation{Center for NanoScience and Department f\"ur Physik,
Ludwig-Maximilians-Universit\"at, Geschwister-Scholl-Platz~1,
80539~M\"unchen, Germany}

\author{A.\,K.~H\"uttel}
\altaffiliation[Present address: ]{Molecular Electronics and Devices, Kavli
Institute of Nanoscience, Delft University of Technology,
PO Box 5046, 2600 GA Delft, The Netherlands}
\affiliation{Center for NanoScience and Department f\"ur Physik,
Ludwig-Maximilians-Universit\"at, Geschwister-Scholl-Platz~1,
80539~M\"unchen, Germany}

\author{K.~Eberl}
\altaffiliation[Present address: ]{Lumics GmbH, Carl--Scheele--Strasse
16, 12489 Berlin, Germany}
\affiliation{Max-Planck-Institut f\"ur Festk\"orperforschung,
Heisenbergstra{\ss}e~1, 70569~Stuttgart, Germany}

\author{S. Ludwig}
\affiliation{Center for NanoScience and Department f\"ur Physik,
Ludwig-Maximilians-Universit\"at, Geschwister-Scholl-Platz~1,
80539~M\"unchen, Germany}

\author{M.\,N.~Kiselev}
\affiliation{Center for NanoScience and Department f\"ur Physik,
Ludwig-Maximilians-Universit\"at, Geschwister-Scholl-Platz~1,
80539~M\"unchen, Germany}
\affiliation{Institut f\"ur Theoretische Physik
I, Universit\"at W\"urzburg, Am Hubland, 97074~W\"urzburg,
Germany}
\affiliation{Material Science Division, Argonne National
Laboratory, Argonne, Illinois 60439, USA}

\author{B.\,L.~Altshuler}
\affiliation{Physics Department, Columbia University, 538 West
120th Street, New York, NY 10027, USA}
\affiliation{NEC-Laboratories America, 4 Independence Way,
Princeton, New Jersey 085540, USA}

\date{\today}


\begin{abstract}
We present transport measurements of the Kondo effect in a double quantum
dot charged with only one or two electrons, respectively. For the one electron
case we observe a surprising quasi-periodic oscillation of the Kondo conductance
as a function of a small perpendicular magnetic field $\left| \bper \right|
\lesssim 50\un{mT}$. We discuss possible explanations of this effect and
interpret it by means of a fine tuning of the energy mismatch of the single dot
levels of the two quantum dots. The observed degree of control implies important
consequences for applications in quantum information processing.
\end{abstract}

\maketitle

The Kondo effect describes a bound state formed by interactions between a
localized magnetic impurity and itinerant conduction band electrons shielding
the localized spin. This results in an increased density of localized states at
the Fermi energy, causing anomalous low temperature properties. In case of a
degenerate ground state of a quantum dot (QD), the Kondo effect manifests itself
as an enhanced conductance within the Coulomb blockade region
\cite{glaz88,ng98,aleiner}. This was first observed on large QDs with half
integer spin \cite{gold98, sara98}, and later, for a total spin of $S=1$, where
the triplet states of a QD are degenerate \cite{schmid99, sasaki00, stopa03}. On
a double quantum dot (DQD) a two-impurity Kondo effect was studied
\cite{jeong01}.

In this article we present the results of Kondo effect differential conductance
(KDC) measurements on a DQD charged with one or two electrons in
a perpendicular magnetic field $B_\bot$. For only one electron ($N=1$) in the DQD
we observe a quasi-periodic structure of the KDC with a characteristic scale of
$B_0\sim 10\,\text{mT}$. In contrast for $N=2$ the KDC is found to be a
monotonic function of $B_\bot$. We discuss possible explanations for this
effect that imply consequences in quantum information processing. 

\begin{figure}[th]\begin{center}
\includegraphics[width=70mm]{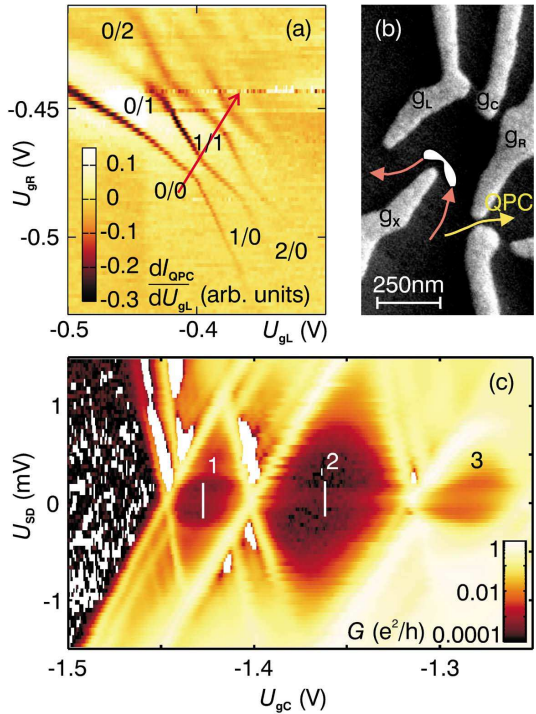}
\vspace*{-6mm}
\caption{\label{kondodiamanten} (Color online)
(a) Stability diagram of the DQD. Plotted is the transconductance
$\mathrm d I_\mathrm{QPC}/\mathrm d U_\mathrm{g_L}$ (color scale) of the DQD as
a function of the side gate voltages $U_\mathrm{g_L}$ and $U_\mathrm{g_R}$. The
measurement was done by using a nearby QPC. A background is subtracted for
clarity. Numerals denote the number of electrons charging the (left/right) QD.
(b) SEM micrograph of the top gates on the sample surface. Arrows mark
possible current paths through the DQD and the nearby QPC. The estimated DQD
geometry is sketched in white.
(c) Differential conductance $\mathrm{d}I/\mathrm{d}U_{\mathrm {SD}}$ of
the symmetrically charged DQD as a function of bias voltage $U_{\mathrm {SD}}$ and
center gate voltage $U_\mathrm{g_C}$. The variation of $U_\mathrm{g_C}$
approximately corresponds to the path in the stability diagram indicated by an
arrow in (a). Numerals indicate the number of electrons charging the DQD.}
\end{center}
\end{figure}

Our sample is fabricated from an AlGaAs/GaAs heterostructure. It embeds a
two-dimensional electron system (2DES) with carrier density
$n_\mathrm{s}\simeq1.8\times 10^{15}\,\mathrm{m^{-2}}$ and electron mobility
$\mu\simeq75\,\mathrm{m^2/Vs}$ (at $T=4.2\,\mathrm K$) $120\,\mathrm{nm}$ below
its surface. Figure \ref{kondodiamanten}(b) shows Ti/Au-gates created by
electron beam lithography. They are used to locally deplete the 2DES to define a
one electron QD.
The gate design is optimized for transport measurements through a QD charged by
only few electrons \cite{ciorga}. By decreasing the voltages applied to
gates $\mathrm{g_C}$ and $\mathrm{g_X}$ (with respect to the 2DES) while
increasing the voltages on the side gates $\mathrm{g_L}$ and $\mathrm{g_R}$ we
deform the QD into a DQD (sketched in Fig.~\ref{kondodiamanten}(b))
\cite{huettel,singledot}. The DQD is tuned to the regime of strong coupling to the
leads and an order of magnitude stronger interdot tunnel coupling of $2t_0\simeq
240\,\mathrm{\mu eV}$ between the adjacent QDs \cite{huettel}. Measurements
are performed in a dilution refrigerator at an electron temperature
$T_\mathrm{2DES}\sim 0.1\,\mathrm{K}$.

A nearby quantum point contact (QPC) is used to detect the charge distribution
of the DQD shown in the stability diagram in Fig.~\ref{kondodiamanten}(a)
\cite{field}. It displays a lock-in measurement of the differential
transconductance $G_\mathrm{QPC}=\mathrm d I_\mathrm{QPC}/\mathrm d
U_\mathrm{g_L}$ as a function of the dc voltages applied to gates $\mathrm{g_L}$
and $\mathrm{g_R}$. In the lower left corner region in
Fig.~\ref{kondodiamanten}(a) the DQD is uncharged (compare figure caption)
\cite{huettel}.

The differential conductance of the DQD is plotted in Fig.~\ref{kondodiamanten}(c)
as a function of the applied bias voltage $U_\mathrm{SD}$ and the center gate
voltage $U_\mathrm{g_C}$. The DQD is tuned such, that the variation in
$U_\mathrm{g_C}$ ($x$-axis) causes a shift in the stability diagram
approximately along the arrow in Fig.~\ref{kondodiamanten}(a). Hence, a charge
between $N=0$ and 3 electrons, marked in Fig.~\ref{kondodiamanten}(c) by numerals,
is distributed symmetrically between the adjacent QDs. Within the diamond-shaped
regions transport is impeded by Coulomb blockade (CB). Nevertheless, strong
coupling of our DQD to its leads allows for inelastic co-tunneling causing an
enhanced conductance within the CB regions at $U_\mathrm{SD}\ne 0$,
(\textit{e.g.} for $N=1$ at $|eU_\mathrm{SD}|\gtrsim 2t_0\simeq
240\,\mathrm{\mu eV}$) \cite{Franceschi}.

In addition, at $U_\mathrm{SD}\simeq 0$ an increased differential conductance is
visible in the CB regions for $N=1,\,2$ or $3$. We assign this zero bias anomaly
to the Kondo effect on a DQD, here charged with only a few electrons. The
observed KDC is small compared to the unitary limit ($G\ll2e^2/h$). This is due
to the tunnel barrier hindering electron transport between the two adjacent QDs
and to an asymmetric coupling to the leads~\cite{huettel}. For $N=1$ or $3$ the
KDC of the DQD can be described by the spin $1/2$ Kondo effect, but for $N=2$
the threefold degenerate triplet states lead to the KDC. This suggests that the
exchange coupling separating the triplet states from the singlet ground state is
smaller than either the Kondo or the electron temperature.

\begin{figure}[th]
\begin{center}
\includegraphics[width=70mm]{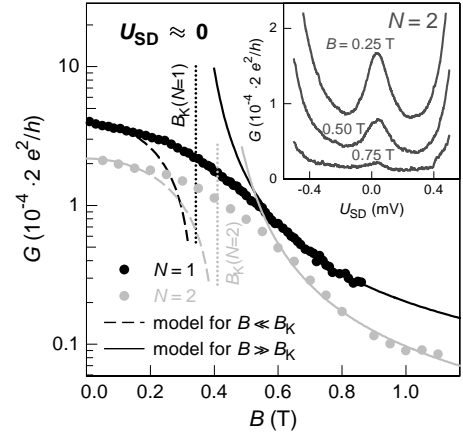}
\vspace*{-8mm} \caption{\label{large_B} Kondo effect
differential conductance in a one (two) electron DQD. The inset
displays exemplary raw data curves of the KDC of the DQD at $N=2$
as a function of the bias voltage for different magnetic fields
but constant gate voltages. All raw data curves are measured
within CB regions as sketched by the vertical white lines in
Fig.~\ref{kondodiamanten}. Black ($N=1$) and grey ($N=2$) circles
in the main figure display the KDC in a logarithmic scale at
$U_\mathrm{SD}\simeq0$ (local maxima of raw data curves) as a
function of a perpendicular magnetic field. Lines are model curves
explained in the main text.}
\end{center}
\end{figure}

Figure~\ref{large_B} displays the KDC at $U_\mathrm{SD}\simeq 0$ of the DQD as a
function of a magnetic field $B_\bot$ perpendicular to the 2DES for $N=1$ and
$N=2$, respectively. Each point corresponds to the maximum KDC near zero bias
measured at constant gate voltage approximately along the white vertical lines
in Fig.~\ref{kondodiamanten}(c). Three of these traces $G(U_\mathrm{SD})$ are
plotted in the inset of Fig.~\ref{large_B}. For increasing $B_\bot$ the KDC is
expected to monotonically decrease as the spin degeneracy is lifted. A
theory by Pustilnik and Glazman provides analytical expressions for the limits
$B\ll B_{\mathrm K}$ and $B\gg B_{\mathrm K}$ \cite{glaz05}, where the
characteristic field $B_{\rm K}$ is determined by the Kondo temperature
$k_\mathrm B T_\mathrm K=g\mu_\mathrm B B_\mathrm K$. For $B_\bot\ll B_{\rm K}$,
the KDC is described by $G\simeq G_0(1-(B_\bot/B_\mathrm K)^2)$ and for
$B_\bot\gg B_{\rm K}$ by $G\simeq{G_\infty} /\,{\ln^2\left(B_\bot/B_\mathrm
K\right)}$. $G_0$ is the KDC at $B_\bot=0$.  The lines in Fig$.$ \ref{large_B}
model these expressions with $T_\mathrm K=0.1\,\mathrm{K}$ for $N=1$ and
$T_\mathrm K=0.12\,\mathrm{K}$ for $N=2$. $T_\mathrm K$ is taken identical for
both limits (solid and dashed lines), respectively. Being close to the
electron temperature of the 2DES, $T_\mathrm K$ cannot be extracted from
temperature dependences as usual. Nevertheless, the
model curves used here are expected to hold even for $T_\mathrm K\sim
T_\mathrm{2DES}$. The agreement with our data is satisfactory.

For $B_\bot\gtrsim 0.5\,\mathrm{T}$ the decrease of the KDC gets steeper due to
a $B_\bot$ dependent decrease of the interdot tunnel coupling, specifically
investigated for our DQD~\cite{huettel}. Taking such effects into account does
not change the Kondo temperature too much compared to the simple model presented
here. However, our simplified model causes the fit-parameter $G_\infty$ to be
strongly suppressed compared to $G_0$.

\begin{figure}[th]
\begin{center}
\vspace*{-5mm}
\includegraphics[width=70mm]{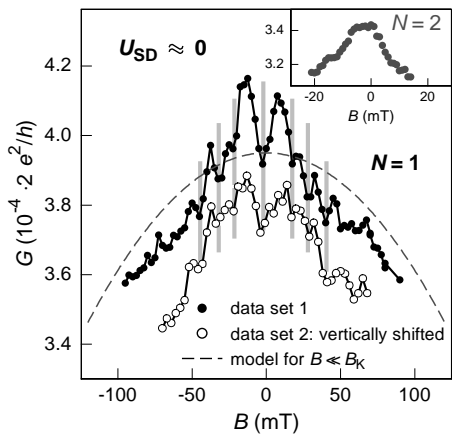}
\vspace*{-5mm}
\caption{\label{small_B}
Non-monotonic magnetic field dependence of the Kondo effect in a one
electron DQD.
Measurements of the KDC as shown in
Fig$.$~\ref{large_B} but for the small magnetic field limit. The
inset and the main figure plot data for $N=2$ and $N=1$,
respectively. One data set (open circles) is vertically shifted
for clarity. The dashed curve describes the limit  $B_\bot\ll
B_{\rm K}$ for $N=1$ and identical parameters as the corresponding
model curve in Fig$.$~\ref{large_B}. Vertical lines mark
local minima of the $N=1$ KDC. Both datasets are taken at
very similar conditions. Nevertheless, in these as in all our measurements
we observe slight differences in the oscillation amplitude.}
\end{center}
\end{figure}

Figure~\ref{small_B} 
shows detailed measurements of the KDC for $N=1$ (main figure) and $N=2$
(inset), as plotted in Fig.~\ref{large_B} but at small
$|B_\bot|<0.1\,\mathrm{T}$. All curves are symmetric in respect to the magnetic
field direction, despite a small offset of $\boffs \simeq -2\,\mathrm{mT}$
caused by a residual background magnetic field. As expected, the KDC for $N=2$
(inset) as well as the co-tunneling differential conductance for $N=1$ (not
shown) decrease monotonically when the magnetic field increases. Surprisingly,
for $N=1$ the KDC shows a non-monotonic behavior. A pronounced local minimum at
$\bper \simeq \boffs \simeq0$ is followed by a quasi-periodic oscillation with
minima at $\left| \bper - \boffs \right| \simeq 0,\,20,\,30,\,42\,\mathrm{mT}$
(vertical lines in Fig$.$~\ref{small_B}). These oscillations quickly decay with
increasing magnetic field and are convolved with the expected decrease of the
KDC for $B\ll B_{\mathrm K}$ (dashed line).

One can consider the following possible explanations for the KDC
observed in a DQD for $N=1$ to be a nonmonotonic function of
$B_\bot$: Namely, (I, II) the leads, (III) nuclear spins, (IV)
Aharonov-Bohm (AB) like interferences, or (V) the alignment of
energy levels of the two adjacent QDs. All but the last of these
possibilities can be ruled out.

(I) Shubnikov-de Haas oscillations in the leads cannot depend on the number of
electrons. Thus, in contrast to our findings they should manifest identically
in both cases for $N=1$ and $2$.

(II) Spin orbit (SO) interaction in the leads may result in a suppression of the
KDC in zero magnetic field due to spin entanglement between the electrons in
the lead and the dot \cite{kikoin}. The SO interaction, however, can not explain
oscillations of the KDC and a quasi-periodic peak structure. Furthermore, the SO
interaction should equally affect the dot whether charged with one or two
electrons. Experimentally, there is no evidence of a conductance minimum for
$N=2$ (see inset of Fig.~\ref{small_B}). We conclude, that the influence of SO
interaction is negligible for the effect we observed.

(III) In GaAs, $\sim 10^5$ nuclear spins form an internal Overhauser field
$B_{\mathrm{nuc}}\sim 10\,\text{mT}$ applied to the electrons on each QD. This
field fluctuates on the time scale of $t_N \sim
10\,\mathrm{ms}$~\cite{khaetskii}. In our lock-in measurements the data are
averaged over a much longer time of $\sim300\,\mathrm{ms}$.  Hence, the
fluctuations of $B_{\mathrm{nuc}}$ are unlikely to be responsible for the
observed oscillations.

(IV) Interference effects in the orbital motion of an electron in
a double well potential, which determines the DQD, could lead to
AB-like oscillations in the amplitude of the tunneling between the
two wells. In terms of the magnetic flux the period of the
oscillation is the flux quantum $\Phi_0= h /e$ \cite{vav05}. The
overlap of the wavefunctions centered in the adjacent wells of a
DQD is proportional to a relative phase shift $\langle
\psi_1|\psi_2\rangle \propto \exp(2\pi i B S_\text b/\Phi_0)$ of
the classically forbidden region $S_\mathrm b\sim Ld$ (we assume a
rectangular barrier of width $d$, lateral extension $L$ and height
$V$ separating the two wells). Therefore, one would expect the
period of the oscillations to be of the order $\Phi_0/S_\mathrm b$
which is $\sim 0.5\,\text T$ for our DQD. This is far in excess of
the typical quasi-period $(\sim 10\,\text{mT})$ of the observed
oscillations (Fig$.$~\ref{small_B}).

As to AB interferences between different tunneling paths, the area enclosed by a
possible AB contour can be estimated to be $S_\mathrm{AB}\sim Ld$ (or even
smaller due to the serial configuration of the DQD).

We also consider the possibility of Fano-like oscillations, which would be
associated with a leakage current under our gates \cite{Fano}. However,
experimentally we exclude leakage currents due to several measurements (not
shown here).

(V) We believe that the observed quasi-periodic oscillations can
be attributed to the magnetic field effect on the alignment of the
energy levels in the two adjacent QDs. There is no reason to
expect that the DQD structure is  perfectly symmetric and the
single electron eigen-energies $\epsilon_{1,2}$ in the two QDs are
exactly identical. The transmission through the barrier
corresponding to a one-dimensional motion of an electron is
determined by the overlap of the wave functions  $ \langle
\psi_1|\psi_2\rangle\sim \sinh(\kappa_-)e^{-\kappa_+}/\kappa_- $,
where $\kappa_\pm=(\kappa_1\pm\kappa_2)/2$ and
$\kappa_{1,2}=\sqrt{2md^2(V-\epsilon_{1,2})}/\hbar$. The tunneling
rate reaches its maximum at resonance for $\kappa_1=\kappa_2$.

Since the eigen-energies and, consequently, the parameters $\kappa_{1,2}$ depend
on the magnetic field, the latter can be used to fine tune the tunneling rate.
What is the magnetic field that can compensate a mismatch
$\Delta\epsilon=|\epsilon^0_1-\epsilon^0_2 |$ between the ground state
eigen-energies of the two wells? Using the 1d Schr\"odinger equation for a
rectangular (or parabolic) double well potential one obtains $\hbar
\omega_{\mathrm c} \sim \sqrt{\epsilon^0_1 \epsilon^0_2}\sqrt{\Delta
\epsilon/W}$, where $W$ is the energy difference between the two local minima of
the double well potential and $\omega_{\mathrm c}= |eB/m|$ is the cyclotron
frequency.

At $\bper=0$ the energy mismatch $\Delta\epsilon$ is finite. It takes about
$B_0\simeq\pm 12\,\mathrm{mT}$ to align the ground states, which corresponds to
the first KDC maximum. The suppression of the mismatch by the magnetic field
explains the pronounced minimum at $\bper=0$. This behavior is obviously
symmetric in respect to the sign of the magnetic field. Note, that the
characteristic magnetic field $B_0$ can be very small due to the factor
$\sqrt{\Delta\epsilon/W}$, roughly estimated to be $\sim 10^{-3}$ for our setup.
The interdot tunneling rate is unaffected by thermal line broadening as long as
it corresponds to low frequency noise allowing adiabatic alignment of the energy
levels in both QDs.

The remaining KDC maxima at slightly larger magnetic fields probably correspond
to alignments of excited energy states in the two QDs. This implies the
importance of co-tunneling processes, which are indeed strong (compare
Fig.~\ref{kondodiamanten}(c)). Moreover, in order to explain the observed
quasi-periodic magnetic field dependence, the level structures in the two QDs
should differ due to some anisotropy. From the number of observed KDC maxima we
conclude that at least three excited states are involved in the co-tunneling.

Our model is consistent with the missing KDC oscillations for the doubly
occupied DQD ($N=2$). Indeed, for $N=2$ and a symmetric charge distribution with
one electron in each dot, transport is determined by the singlet and triplet
states. The symmetric charge distribution allows enhanced elastic co-tunneling
decreasing the dependence on the misalignment of the single dot ground states.

In conclusion, we here presented measurements of the Kondo effect on a DQD
charged by only one or two electrons. We demonstrate control of
the resonant tunneling in the one electron case by means of a magnetic field
which appears to be surprisingly small. A non-monotonic magnetic field
dependence of the KDC is attributed to the anisotropy of the DQD. The magnetic
field fine tunes the alignment of energy levels in the adjacent QDs, modifying
the interdot tunnel splitting. Hence, the magnetic field provides an extremely
sensitive tool to detect and control the anisotropy of a single electron DQD.

Our double dot with strong coupling to the leads is not a perfect system for a
qubit. However, the influence of a very small misalignment of the single dot
energy levels onto the tunnel probability of the electron will remain even for a
double dot weakly coupled to the leads, which can be used e.g.\ as a charge
qubit \cite{hayashi,wiel}. The observed variations of the interdot tunnel splitting at very small
magnetic fields have to be taken into account for the design of 
semiconductor nano-devices in the field of quantum information processing
\cite{lossdivi}.

We thank Yu.\ Galperin, L.\ I.\ Glazman, M.\ P.\ Pustilnik, A.\
O.\ Govorov, J.\ von Delft, K.Kikoin and J.\ Kotthaus for helpful
discussions. We acknowledge financial support by the BMBF via
DIP-H2.1, the DFG via the SFB 631. MK acknowledges support through
the Heisenberg program of the DFG. Research in Argonne was
supported by U.S. DOE, Office of Science, under Contract No.
W-31-109-ENG-39.

\bibliographystyle{apsrev}

\end{document}